\documentstyle[11pt,paspconf,graphics]{article}

\begin{document}

\title{The Growth of Dark-Matter Halos. The Effects of Accretion and Mergers}

\author{E. Salvador-Sol\'e\altaffilmark{1}, and A. Manrique\altaffilmark{2}}
\affil{Department d'Astronomia i Meteorologia, Universitat de Barcelona,
    Barcelona, E-08028, Spain}

\author{J.M. Solanes\altaffilmark{3}}
\affil{Departament d'Enginyeria Inform\`atica, Universitat Rovira i
Virgili, Tarragona, E-43006, Spain}

\altaffiltext{1}{eduard@faess0.am.ub.es} 
\altaffiltext{2}{alberto@pcess2.am.ub.es} 
\altaffiltext{3}{solanes@jsolanes.etse.urv.es}

\begin{abstract}
High resolution cosmological N-body simulations show that the density
profiles of dark matter halos in hierarchical cosmogonies are
universal, with low mass halos typically denser than more massive
ones. This mass-density correlation is interpreted as reflecting the
earlier formation of less massive objects. We investigate this
hypothesis in the light of formation times defined as the epoch at
which halos experience their last major merger. We find that the
characteristic density and the scale radius of halos are essentially
proportional, respectively, to the critical density of the universe and
the virial radius at the time of their formation. These two relations
are consistent with the following simple evolutionary picture. Violent
relaxation caused by major mergers rearrange the structure of halos
leading to a universal dimensionless density profile. Between major
mergers, halos gradually grow through the accretion of surrounding
layers by keeping the central part steady and only expanding their
virial radius as the critical density of the universe diminishes.
\end{abstract}

\keywords{cosmology,galaxy formation,clusters of galaxies,dark-matter}

%
%
\def\roc{\rho_{\rm c}}
\def\rs{r_{\rm s}}
\def\delc{\delta_{\rm c}}
\def\xs{x_{\rm s}}
\def\ixs{\xs^{-1}}
\def\delf{\delta_{\rm cf}}
\def\xsf{x_{\rm sf}}
\def\tf{t_{\rm f}}
\def\delm{\Delta_{\rm m}}
\def\xnfw{\xsf^{{}^{\rm (NFW)}}}
\newcommand{\der}{{\rm d}}

\section{Introduction}

Despite the big efforts made in the last two decades in trying to
understand the origin of the density profile of galaxies and clusters,
the situation is still very confused. Is the typical density profile of
halos mainly the result of accretion (spherical secondary infall) or of
repetitive major mergers (strong violent relaxation)? Does it arise, on
the contrary, from secular evolution due to the effects of dynamical
friction plus the tidal stripping of captured satellites? Or is it some
combination of the preceding processes? 

Recently, two empirical facts have emerged from high resolution
cosmological N-body simulations which seem to have important
implications on this longstanding debate. First, the spherically
averaged density profile of dark-matter halos is universal (Navarro,
Frenk, \& White 1997, hereafter NFW; Cole \& Lacey 1997; Tormen, Bouchet,
\& White 1997), its exact form being however controversial (see
Moore et al. 1998). The various laws proposed can be encapsulated
within the general expression
\begin{equation}
{\rho(\xi)\over{\rho_{\rm crit}}}={\delta_{\rm c} \over
\xi^{\alpha}(1+\xi^{\beta})^{(\gamma-\alpha)/\beta}}\;,
\label{rho}           
\end{equation}
with $\alpha=1$, $\beta=1$, and $\gamma=3$ for the NFW profile, and
$\alpha=1.4$, $\beta=1.4$, and $\gamma=2.8$ for the Moore et al.~(1998)
profile. The classical King law and the Hernquist~(1990)
profile also admit this general form for $\alpha=0$, $\beta=2$, and
$\gamma=3$, and $\alpha=1$, $\beta=1$, and $\gamma=4$, respectively. In
equation (\ref{rho}), $\xi=r/\rs$ is the radial distance to the halo
center in units of the scale radius $\rs$, and $\delc=\roc/\rho_{\rm
crit}$ is the characteristic halo density in units of the critical
density of the universe. The dimensional parameters $\roc$ and $\rs$
are linked by the condition that the mean density within the virial
radius $R$ of a halo of a given mass is a constant factor $a$ times
$\rho_{\rm crit}$; here we adopt $a=200$. Therefore, the density
profiles of halos at a given epoch depend on their mass $M$ through a
unique parameter, $\delc$ or $\xs=\rs/R$. Second, the smaller the mass
of halos, the denser they are (NFW, Cole \& Lacey 1997) or,
equivalently, $\delc$ is a decreasing function of $M$.

This mass-density correlation is interpreted as reflecting the fact
that, in hierarchical clustering, less massive halos form typically
earlier when the mean density of the universe is higher. According to
this interpretation the structural properties of halos would be fixed
at their time of formation. As major mergers yield a substantial
rearrangement of the system, while accretion only causes its smooth
evolution, this would point to the fact that violent relaxation taking
place after major mergers is the main responsible for the halo density
profiles.  NFW have shown, indeed, that the characteristic density of
halos with a given mass at a given epoch is proportional to the mean
cosmic density when they form. This connection between the
characteristic density of halos and their formation time strongly
favors the previous interpretation of the mass-density
correlation. However, the physics behind such a proportionality is
unclear. Moreover, the halo formation time adopted by NFW does not
correspond to the last major merger experienced by these systems.

Here we report the results of a similar analysis of the mass-density
correlation recently performed by Salvador-Sol\'e, Solanes, \& Manrique
(1998, SSM) using a better suited formation time estimate. Not only
does our analysis confirm the origin proposed by NFW for the
mass-density correlation, but also provides a physical base for the
proportionality between halo characteristic density and cosmic density
at formation. Furthermore, our results point to a complete picture
for the evolution of halo structure as a consequence of the interplay
of accretion and major mergers.

\section{A Better Suited Halo Formation Time Estimate}

To follow the formation and evolution of halos, SSM have used a
slightly modified version of the extended Press-Schechter (1974, PS)
clustering model, known to agree with N-body simulations. The only
change introduced consists on including a schematic distinction between
minor and major mergers. Such a distinction does not obviously alter
the success of the basic model while it allows one to define the
formation of a halo as the last major merger it experiences.

We say that a halo of mass $M$ experiences a major merger and is
destroyed when the relative mass captured by a halo $\Delta M/M$, with
$\Delta M=M'-M$ the increment of mass, exceeds a certain threshold
$\delm$. Otherwise, the event is regarded as an accretion and the
capturing halo keeps its identity. Consequently, the mass-accretion and
destruction rates of halos of mass $M$ at $t$ are respectively given by
\begin{equation}
r^{\rm a}_{\rm mass}(M,t)=\int_M^{M(\Delta_{\rm m}+1)}\,\Delta
M\,r^{\rm m}_{\rm LC}(M\rightarrow M',t)\,\der M'
\label{accm}
\end{equation}
and
\begin{equation}
r^{\rm d}(M,t)=\int_{M(\Delta_{\rm m}+1)}^\infty r^{\rm m}_{\rm LC}
(M\rightarrow M',t)\,\der M'\, ,
\label{dr}
\end{equation}
where $r^{\rm m}_{\rm LC}(M\rightarrow M',t)$ is the specific merger
rate in the usual extended PS model (Lacey \& Cole 1993). 

The distinction between minor and major mergers does not modify the
mass function of halos at $t$, $N(M,t)$, which therefore takes the
standard PS form. 

As a result of major mergers, some halos are destroyed and other halos
form. The formation rate of halos with mass $M=M(t)$ can then be
obtained from the conservation equation
\begin{equation}
r^{\rm f}[M(t),t]={\der\ln N[M(t),t]\over \der t}+r^{\rm d}[M(t),t]+
\partial_M r^{\rm a}_{\rm mass}(M,t)\bigl|_{M=M(t)} \label{conserv}
\end{equation}
for the number density of halos per unit mass along mean accretion
tracks, $M(t)$, solution of the differential equation $\der M/\der
t=r^{\rm a}_{\rm mass}[M(t),t]$.

Finally, from this formation rate, one can compute the distribution of
formation times for halos of mass $M_0$ at $t_0>t$ (see SSM for details)
\begin{equation}
\Phi_{\rm f}(t)=r^{\rm f}[M(t),t]\,
\exp\biggl\{-\int_t^{t_0} r^{\rm f}[M(t'),t']\,\der t'\biggr\}\, ,
\label{dtf}
\end{equation}
with $M(t)$ the mean accretion track satisfying $M(t_0)=M_0$. The
median value for this distribution function is then taken as the
typical formation time $\tf(M_0,t_0)$ of such halos.

\begin{figure}
  \resizebox{\hsize}{!}{\includegraphics{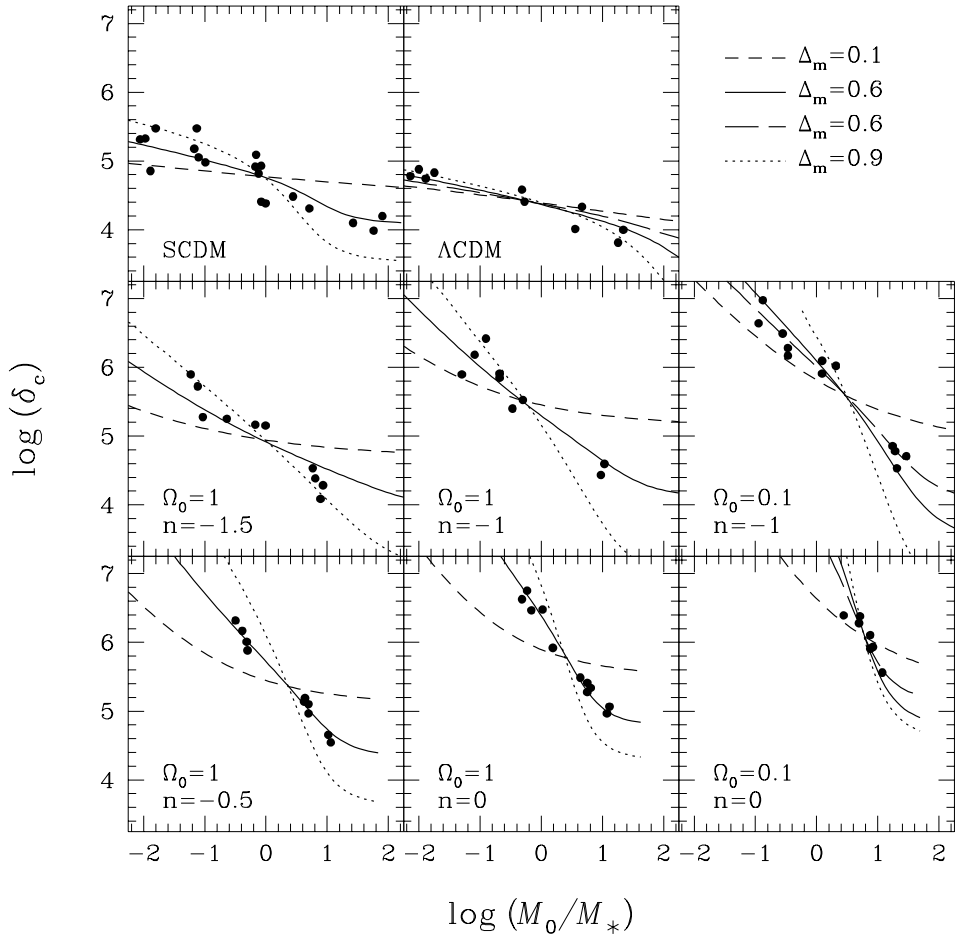}}
\caption{Predicted $\delta_{\rm c}(M_0)$ correlations compared with the
empirical data from NFW's N-body simulations (filled circles). Dotted
and short-dashed curves show the predictions for two extreme values of
$\Delta_{\rm m}$, while the solid curves correspond to the value of
this parameter that gives the best overall fit. Cosmogonies with
$\Omega_0<1$ contain a fourth long-dashed curve which shows, for
$\Delta_{\rm m}=0.6$, the predictions arising from the assumption that
$\rho_{\rm c}$ is proportional to $\rho_{\rm crit}[z_{\rm f}(M_0)]$,
instead of to the mean density of the universe at halo formation.}
\label{fig-1}
\end{figure}

At this stage, the value of $\delm$ establishing in a simple way the
frontier between accretion and major mergers must be regarded as a
phenomenological parameter of the model. Its value will be fixed in
the next section through fits to the empirical mass-density (or
mass-radius) correlation, the only data which seems to be connected
with the distinction between merger and accretion.

\section{The Mass-Density and Mass-Radius Correlations}
As shown in Figure \ref{fig-1}, the empirical $\delta_{\rm c}(M_0)$
correlation for present halos is well fitted, in all the cosmogonies
analyzed, by the simple proportionality between the halo characteristic
density $\rho_{\rm c}$ and the mean density $\bar\rho(z_{\rm f})$ of
the universe when halos form proposed by NFW
\begin{equation}
\delta_{\rm c}=C\Omega_0[1+z_{\rm f}(M_0)]^3,\label{fit}
\end{equation}
provided that the mass threshold for merger takes the very natural value of
0.6.

Similarly good fits are also obtained for this same value of $\delm$ in
low $\Omega$ cosmogonies if $\rho_{\rm c}$ is taken instead
proportional to the critical density for closure $\rho_{\rm cir}(z_{\rm
f})$ at the time of halo formation. This new model, which assumes that
the dimensionless characteristic density of halos at their time of
formation is equal to the cosmogony-dependent constant $\delta_{\rm
cf}$, while the value of its dimensional counterpart $\roc$ remains
fixed since that epoch, is described by the relation
\begin{equation}
\delta_{\rm c}=\delta_{\rm cf}{\Omega_0\over
\Omega[z_{\rm f}(M_0)]}\,[1+z_{\rm f} (M_0)]^3.\label{fit2}
\end{equation}

In fact, the similarity between the empirical distribution function of
halo characteristic densities and the theoretical one implied by this
latter relation, suggests that eq.~(\ref{fit2}) holds also for
individual halos. More importantly, in contrast with the relation
(\ref{fit}) proposed by NFW, eq.~(\ref{fit2}) has a clear physical
interpretation: violent relaxation occurring after major mergers produces
virialized halos with identical dimensionless density profiles, which
remain essentially unaltered during the accretion phase, except for the
continuous stretching of their outer boundaries caused by the secular
decrement of the critical density of the universe.

\begin{figure}
  \resizebox{\hsize}{!}{\includegraphics{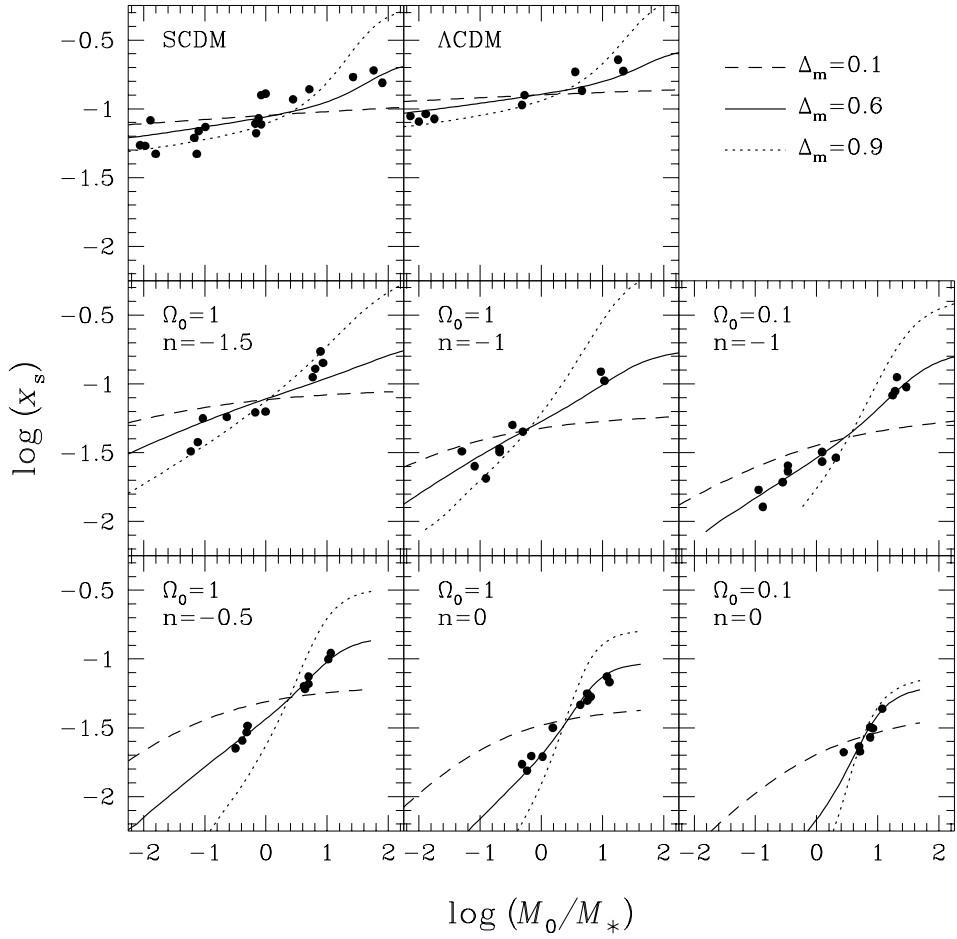}}
\caption{Predicted $x_{\rm s}(M_0)$ correlations compared with the empirical
data from NFW's N-body simulations (solid circles).}
\label{fig-2} 
\end{figure}

Given the tight relation between the characteristic density and the
scale radius of halos of a given mass at a given epoch, if the
evolutionary scheme we are proposing is correct, the value at formation
of the dimensionless scale radius $\xs$ of halos should also be
universal, while the associated dimensional parameter $\rs$ should keep
its value invariant between major mergers. Under these circumstances,
the scale radius $\rs$ for current halos of mass $M_0$ should be
proportional to the virial radius of the halos when they formed, which
implies that the empirical mass-radius correlation associated with the
mass-density one should be well fitted by the expression
\begin{equation}
x_{\rm s}=x_{\rm sf}{R\{M[z_{\rm f}(M_0),M_0]\}\over
R(M_0)}.\label{fit3}
\end{equation}
We want to stress that this is a new relation, independent from the two
previous ones (eqs.~[\ref{fit}] and [\ref{fit2}]). Indeed, to verify it
we need to estimate the mass of halos at the time of their formation
(i.e., when they experienced their last major merger), so a clustering
model such as the one developed in the preceding section is
necessary. In contrast, the verification of the two previous relations
does not require any clustering model.

As can be seen from Figure \ref{fig-2}, relation (\ref{fit3}) gives
very good fits, indeed, to the empirical correlations for all
cosmogonies, provided, once again, that $\delm$ is equal to
0.6. Furthermore, as shown in Table \ref{tbl-1}, the best fitting
values of $\xsf$ are in very good agreement with the values of this
parameter implied independently by the best fits obtained from relation
(\ref{fit2}) above. These results therefore give strong support our
physical interpretation of the fitting formulae.

\begin{table}
\caption{Dimensionless parameters in various Cosmogonies}\label{tbl-1}
\begin{center}
\begin{tabular}{crrrcccc}
\tableline

$P(k)$ & $\Omega_0$ & $\lambda_0$  & $\sigma_8$ & $C$ &
$\delta_{\rm cf}$   & $x_{\rm sf}$\tablenotemark{a}  & 
$x_{\rm sf}$\tablenotemark{b} \\


SCDM & 1.0 & 0.0 & 0.63 & 1.21$\times 10^4$ & 
	1.21$\times 10^4$ & 0.173 & 0.229 \\
$\Lambda$CDM & 0.25 & 0.75 & 1.3 & 4.21$\times
	10^3$ & 3.77$\times 10^3$ & 0.291 & 0.285 \\
$n=-1.5$ & 1.0 & 0.0 & 1.0 & 8.30$\times 10^3$ &
	8.30$\times 10^3$ & 0.204 & 0.223 \\
$n=-1.0$ & 1.0 & 0.0 & 1.0 & 1.28$\times 10^4$ &
	1.28$\times 10^4$ & 0.169 & 0.181 \\
  & 0.1 & 0.0 & 1.0 & 2.65$\times 10^4$ &
	1.00$\times 10^4$ & 0.188 & 0.184 \\
$n=-0.5$ & 1.0 & 0.0 & 1.0 & 2.16$\times 10^4$ &
	2.16$\times 10^4$ & 0.135 & 0.148 \\
$n= 0.0$ & 1.0 & 0.0 & 1.0 & 6.19$\times 10^4$ &
	6.19$\times 10^4$ & 0.088 & 0.096 \\
  & 0.1 & 0.0 & 1.0 & 5.77$\times 10^5$ & 
	1.33$\times 10^5$ & 0.064 & 0.065 \cr
\tableline

\end{tabular}
\tablenotetext{a}{implied by $\delta_{\rm cf}$}
\tablenotetext{b}{from direct fits to the empirical mass-radius relation}
\end{center}
\end{table}

\section{Summary and Discussion}

For $\Delta_{\rm m}$ $\sim 0.6$, the empirical mass-density obtained in
a number of different cosmogonies is very well fitted by the simple
relations (\ref{fit}) or (\ref{fit2}). This confirms, with a physically
motivated halo formation time estimate, the previous claim by NFW that
the characteristic density shown by dark halos in equilibrium is
proportional to the mean or critical density of the universe at the
time they form. A relation like eq.~(\ref{fit2}) is naturally expected
in the case that halos show a universal dimensionless density profile
at the time of formation, and then keep the central part steady while
the virial radius expands as the density of the universe (and its
critical value) diminishes. This simple interpretation of the
mass-density correlation is supported by the fact that relation
(\ref{fit3}) also gives, for $\Delta_{\rm m}$ $\sim 0.6$, very good
fits to the empirical mass-radius correlation, with values of $\xsf$
fully consistent with those of $\delf$ found in the previous fit.

Such a behavior for the scaling parameters of halo density profiles is
very well understood as the combined result of the different dynamical
effects of major mergers and accretion. The strong violent relaxation
due to the former would produce an important rearrangement of the halo
structure leading to essentially a universal dimensionless density
profile only dependent on the particular cosmogony considered. On the
contrary, the smooth growth via spherical secondary infall due to
accretion would keep the halo central body unaltered and gradually
build an extended envelope following the slow decrease of the mean
density of the universe. 

In another talk in this meeting (see Gonz\'alez-Casado, Raig, \&
Salvador-Sol\'e 1998) we report the results of the detailed check of
such an evolutionary scheme for halos. As it is shown there, the values
of parameters $\delf$ or $\xsf$ are completely fixed in such a
scenario, and it turns out that they are consistent with the values
found (see Table \ref{tbl-1}) in the fits to the empirical mass-density
and/or mass-radius correlations.

\acknowledgments
This work has been supported by the Direcci\'on
General de Investigaci\'on Cient\'\i fica y T\'ecnica under contract
PB96-0173.

\end{document}